\documentclass[pra,reprint,superscriptaddress]{revtex4-1}
\usepackage{amsmath}
\usepackage{amssymb}
\usepackage{graphicx}
\usepackage[caption=false]{subfig}
\usepackage{enumitem}
\usepackage{color}
\usepackage[percent]{overpic}
\usepackage{bm}
\usepackage{rotating}
\usepackage{hyperref}
\usepackage{pbox}
\usepackage{array}
\usepackage[normalem]{ulem} 
\usepackage[normalem]{ulem}

\newcommand{\ii}{\mathrm{i}}

\begin{document}

\title{Untangling entanglement and chaos}
\author{Meenu Kumari}
\affiliation{Institute for Quantum Computing, University of Waterloo, Waterloo, Ontario, N2L 3G1, Canada}
\affiliation{Department of Physics and Astronomy, University of Waterloo, Waterloo, Ontario, N2L 3G1, Canada}
\author{Shohini Ghose}
\affiliation{Department of Physics and Computer Science, Wilfrid Laurier University, Waterloo, Ontario, N2L 3C5, Canada}
\affiliation{Institute for Quantum Computing, University of Waterloo, Waterloo, Ontario, N2L 3G1, Canada}
\affiliation{Perimeter Institute for Theoretical Physics, 31 Caroline St N, Waterloo, Ontario, N2L 2Y5, Canada}

\begin{abstract}
We present a method to calculate an upper bound on the generation of entanglement in any spin system using the Fannes-Audenaert inequality for the von Neumann entropy. Our method not only is useful for efficiently estimating  entanglement, but also shows that entanglement generation depends on the distance of the quantum states of the system from corresponding minimum-uncertainty spin coherent states (SCSs). We illustrate our method using a quantum kicked top model, and show that our upper bound is a very good estimator for entanglement generated in both regular and chaotic regions. In a deep quantum regime, the upper bound on entanglement can be high in both regular and chaotic regions, while in the semiclassical regime, the bound is higher in chaotic regions where the quantum states diverge from the corresponding SCSs. Our analysis thus explains previous studies and clarifies the relationship between chaos and entanglement.   
\end{abstract}
\maketitle
\section{Introduction}
Classical chaos is a well-defined property of nonlinear deterministic dynamical systems and is characterized by exponential instability due to sensitivity to initial conditions \cite{ott2002chaos}. On the other hand, entanglement is a property of quantum systems \cite{nielsen2010quantum}. The relationship between the two has intrigued physicists for a couple of decades \cite{Furuya1998,Miller1999,Arul2001,Furuya2001,Arul2002,Arul2003,Fujisaki2003,Jacquod2004,Arul2004,Santos2004,Bambi2004,entanglement2004PRE,entanglement2004PRA,Ghose2005,NOVAES2005308,Prosen2005,Furuya2005,Calsamiglia2005,Lorenza2006,Jacquod2006,Arul2007,Lorenza2008,Chaos2008,Vaibhav2008,chaudhury2009nature,Chung2009,Angelo2010,lombardi2011,Angelo2012,ernest2016,neill2016ergodic,valdez2016many,xu2017entanglement,pattanayak2017,Lewenstein2018quantum,dogra2018quantum,MADHOK2018189,krithika2018nmr,Udaysinh2018}. It has been explored mostly in the semiclassical regime through studies of various models such as the N-atom Jaynes-Cummings model \cite{Furuya1998,Furuya2001}, kicked coupled tops \cite{Miller1999,Arul2004,Vaibhav2008} and the quantum kicked top \cite{entanglement2004PRE,entanglement2004PRA,Lorenza2006,Chaos2008,chaudhury2009nature,lombardi2011,pattanayak2017,Lewenstein2018quantum}. An intriguing aspect is the observation of signatures of chaos in a deep quantum regime where quantum-classical correspondence is not usually expected.  \cite{chaudhury2009nature,neill2016ergodic}. Additionally, the universality of the relationship between chaos and entanglement remains a topic of  debate \cite{lombardi2011,lombardi2015,Vaibhav2015,pattanayak2017} that needs resolution. These studies are of interest not only for a fundamental understanding of quantum-classical correspondence, but also for applications in  quantum computing where entanglement is an important resource \cite{Shepelyansky2000,Shepelyansky2001PRL,Shepelyansky2001PRL2,Shepelyansky2001,Boixo2018}.

Whereas chaotic systems can typically generate large entanglement, some regular systems can also produce high entanglement. In this paper, we explain the puzzling connection between chaos and entanglement generation in spin systems. We provide a framework to determine an upper bound on the entanglement dynamics in any spin system with a constant spin value $j$ (symmetric multiqubit systems). Our framework helps to identify when the bound will remain low and limit the entanglement generated. We show that this is the case when the quantum states remain  close to minimum-uncertainty classical-like spin coherent states (SCSs). The bound grows as the distance between the quantum and classical states increases. Thus entanglement is associated with non-classical dynamics and the breakdown of quantum-classical correspondence. This applies to both regular and chaotic systems. 

We illustrate our framework and upper bound in a model quantum kicked top (QKT) system. We show that our bound provides a very good estimate of the entanglement generated in both regular and chaotic regions. We also analyze regular versus chaotic dynamics in the deep quantum and semiclassical regime, and show that entanglement remains low in regular regions only in a semiclassical regime. Our analysis resolves previous debates about the relationship between entanglement and chaos in a deep quantum versus semiclassical regime. Furthermore, our framework can be used to obtain a computationally efficient loose bound on entanglement. 

\section{Finding an upper bound on entanglement} 
Our method for establishing an upper bound on entanglement relies on the Fannes inequality for von Neumann entropy \cite{Fannes1973}. In \cite{audenaert2007}, Audenaert presented a sharper version of the original Fannes inequality. Consider two quantum states, $\rho$ and $\sigma$, which belong to Hilbert space of dimension $d < \infty$. Let the trace distance between the states be $D = \frac{1}{2} ||\rho - \sigma ||_1$ \cite{nielsen2010quantum}. Then, the difference between the von Neumann entropy of the two states is bounded by
\begin{equation}
|S(\rho) - S(\sigma)| \leq D\text{log}_2(d-1)+h(D)
\label{Fannes}
\end{equation}
where $h(D)=-D\text{log}_2(D)-(1-D)\text{log}_2(1-D)$. 

Our focus is on spin systems having constant angular momentum $j$, such as Dicke models \cite{Dicke1954}, the QKT \cite{haake1987}, and Lipkin-Meshkov-Glick (LMG) model Hamiltonians \cite{Vidal2004} that have no coupling to other fields. The states of such systems lie in the symmetric subspace of $N=2j$ spin-1/2 qubits. Thus one can study entanglement by considering any bipartition of the $N$ qubits, say the $m{:}(N-m)$ bipartition where $m\in \{1,..N-1\}$. A minimum uncertainty state corresponds to a spin coherent state (SCS), $|j, \Theta,\Phi \rangle = \exp{[\ii \theta(J_x \sin{\phi} - J_y \cos{\phi})]}|j,j\rangle$. In the multiqubit representation, $\text{SCS} |j, \Theta,\Phi \rangle = |\theta,\phi \rangle \otimes |\theta,\phi \rangle \otimes ...... \otimes |\theta,\phi \rangle \text{     (}2j\text{ times})$, where $|\theta, \phi \rangle = \cos{\left( \frac{\theta}{2} \right)}|0\rangle + \exp{(\ii \phi)}\sin{\left( \frac{\theta}{2} \right)}|1\rangle$.

Consider an initial spin quantum state that has evolved after a time $t$ to the state $|\psi (t) \rangle$. Let $D$ denote the trace distance between $|\psi (t) \rangle$ and the SCS $|j, \Theta_{\text{ev}},\Phi_{\text{ev}} \rangle$ centered at $(\Theta_{\text{ev}},\Phi_{\text{ev}})$ corresponding to the expectation values of $|\psi (t) \rangle$, with $(j\sin{\Theta_{\text{ev}}}\cos{\Phi_{\text{ev}}},j \sin{\Theta_{\text{ev}}}\sin{\Phi_{\text{ev}}},j \cos{\Theta_{\text{ev}}}) = (\langle J_x \rangle, \langle J_y \rangle, \langle J_z \rangle)$. Let $D_{\text{re}}$ be the trace distance between the $m$-qubit reduced states of $|\psi (t) \rangle$ and SCS$|j, \Theta_{\text{ev}},\Phi_{\text{ev}} \rangle$, that is, $\rho_m(t)$ and $\rho_m^{SCS}$ respectively. Then $D_{\text{re}} \leq D$ since trace distance is non-increasing under partial trace \cite{nielsen2010quantum}. The $m{:}(N-m)$ entanglement in $|\psi(t) \rangle$ is quantified by $S(\rho_m(t))$. The SCS is a product state and thus, $S(\rho_m^{\text{SCS}}) = 0$. Using these in \eqref{Fannes}, we obtain the following bound on $m{:}(N-m)$ entanglement in $|\psi(t) \rangle$:
\begin{equation}
    S(\rho_m(t))  \leq  D_{\text{re}}\text{log}_2(d-1)+h(D_{\text{re}}).
        \label{bound1}
\end{equation}
Here, $d$ is the dimension of $\rho_m(t)$ and $\rho_m^{\text{SCS}}$ (which can be less than $2^m$ owing to the symmetries of the state). This bound on entanglement in \eqref{bound1} demonstrates that the generation of entanglement in any system depends on the trace distance between relevant states. When the state remains close to the minimum uncertainty classical-like SCS, the entanglement remains low. As the trace distance from these SCS grows, the second and higher-order cumulants in the evolving state grow, leading to a divergence from classicality. This illustrates how a divergence from classicality is associated with the generation of entanglement. In the case of chaotic systems, the states do not remain close to SCS, even in the semiclassical regime. Thus, the bound goes to a maximum for chaotic systems. 

In our framework, one can choose a different SCS to compute the trace distance, which will yield a different $D_{\text{re}}$, and a different and potentially tighter upper bound for $S(\rho_m(t))$. Consider the SCS corresponding to a state which is obtained from the classical evolution of the initial state for time $t$, namely, $|j, \Theta_{\text{cl}},\Phi_{\text{cl}} \rangle$, corresponding to the classical state $(J_x(t),J_y(t),J_z(t))$, with $(j\sin{\Theta_{\text{cl}}}\cos{\Phi_{\text{cl}}},j \sin{\Theta_{\text{cl}}}\sin{\Phi_{\text{cl}}},j \cos{\Theta_{\text{cl}}}) = (J_x(t),J_y(t),J_z(t))$. Let $D_{\text{re}}^{\prime}$ denote the trace distance between $\rho_m(t)$ and the $m$-qubit reduced state of the SCS $|j, \Theta_{\text{cl}},\Phi_{\text{cl}} \rangle$. Then $D_{\text{re}}$ in \eqref{bound1} can be taken as the minimum of $D_{\text{re}}^{\prime}$ and the $D_{\text{re}}$ described in the previous paragraph. Thus, \eqref{bound1} becomes

\begin{equation}
    S(\rho_m(t))  \leq  \text{min}(D_{\text{re}},D_{\text{re}}^{\prime})\text{log}_2(d-1)+h(\text{min}(D_{\text{re}},D_{\text{re}}^{\prime})).
        \label{bound2}
\end{equation}
This can yield a better bound on entanglement as we will illustrate in the QKT model described below. One can understand the physical motivation of choosing the two SCSs described above from the Ehrenfest correspondence principle, which examines when the expectation values of observables obey the classical equation of motion. One of the SCSs is constructed from the expectation value of observables, and the second SCS is constructed from the classical equation of motion. Interestingly, the relationship between entanglement and distance from SCSs is similar to the dependence of generalized entanglement on the spread of quantum states~\cite{Lorenza2006}.

We can further use this framework to obtain a computationally efficient but loose bound to the entanglement. As already mentioned, $D_{\text{re}}\leq D$. Second, the RHS in \eqref{Fannes} is a monotonically increasing function of $D$ up to $D=(1-1/d)$ where $D \in [0,1]$ \cite{audenaert2007}. Thus, for $D\leq 1-1/d$, $D\text{log}_2(d-1)+h(D)$ will serve as a loose bound on entanglement in \eqref{bound1}. This loose bound is numerically inexpensive to calculate for any $m:(2j-m)$ bipartition in comparison to the von Neumann entropy and the RHS bound in \eqref{bound1} \cite{pan1999complexity,maziero2017computing}.

\section{Entanglement in the QKT}

\begin{figure}
\centering\includegraphics[width=0.5\textwidth]{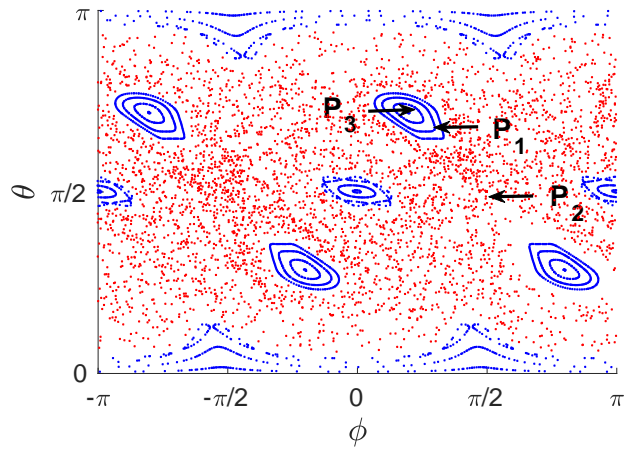}
\caption{[Color online] Classical stroboscopic map for the kicked top showing a mixed phase space of regular (blue) islands in a chaotic (red) sea.  $P_1$, $P_2$ and $P_3$ are 3 points corresponding to $(\theta,\phi) = (2.1,0.9), (1.5,1.5)$ and $(2.25,0.75)$ respectively. $\kappa=3$, $p=\pi/2$, and $\tau=1$.} 
\label{Classical}
\end{figure}

\begin{figure*}
\centering\includegraphics[width=\textwidth]{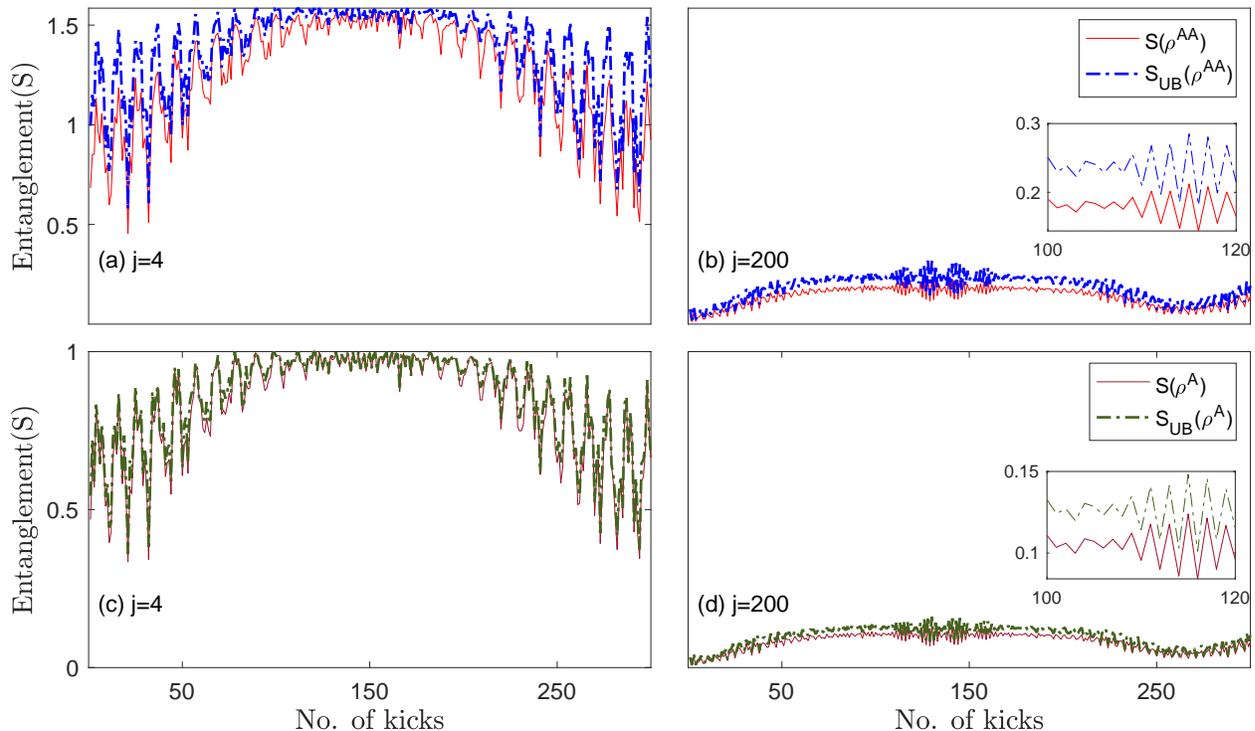}
\caption{Evolution of QKT entanglement and the upper bound in \eqref{bound2} for an initial state in the regular region, $P_1$ in Fig. \ref{Classical}. (a) $2{:}(N-2)$ entanglement for $j=4$, (b) $2{:}(N-2)$ entanglement for $j=200$, (c) $1{:}(N-1)$ entanglement for $j=4$, and (d) $1{:}(N-1)$ entanglement for $j=200$. The upper bound on entanglement is almost saturated by the entanglement generated in the kicked top.} 
\label{Regular}
\end{figure*}

The quantum kicked top (QKT) is a multiqubit, time-periodic spin system whose classical counterpart exhibits a plethora of interesting features, such as a mixed phase space, bifurcations, and chaos \cite{haake1987,kumari2018quantum}. The QKT Hamiltonian is
\begin{equation}
H = \hbar \frac{\kappa}{2 j \tau} J_z^2 + \hbar p J_y \sum_{n= - \infty}^{\infty} \delta (t - n \tau),
\label{top1}
\end{equation}
where $\kappa$ and $p$ are parameters, $J_x$, $J_y$ and $J_z$ are angular momentum operators, and $j$ is a constant of motion since $J^2$ commutes with the Hamiltonian. The unitary operator for the kicked top corresponding to one time period, $\tau$, is given by $U=\exp{(-i \frac{\kappa}{2j}J_z^2)} \exp{(-ipJ_y)}$. The classical equations of motion for the kicked top can be obtained by writing the Heisenberg equation of motion for $J_x/j$, $J_y/j$, and $J_z/j$, and taking the limit $j \rightarrow \infty$. Figure \ref{Classical} shows the classical stroboscopic map of the kicked top. It shows a mixed phase space of regular islands surrounded by a sea of chaotic dynamics for $\kappa=3$ and $p=\pi/2$. 

\begin{figure*}
\centering\includegraphics[width=\textwidth]{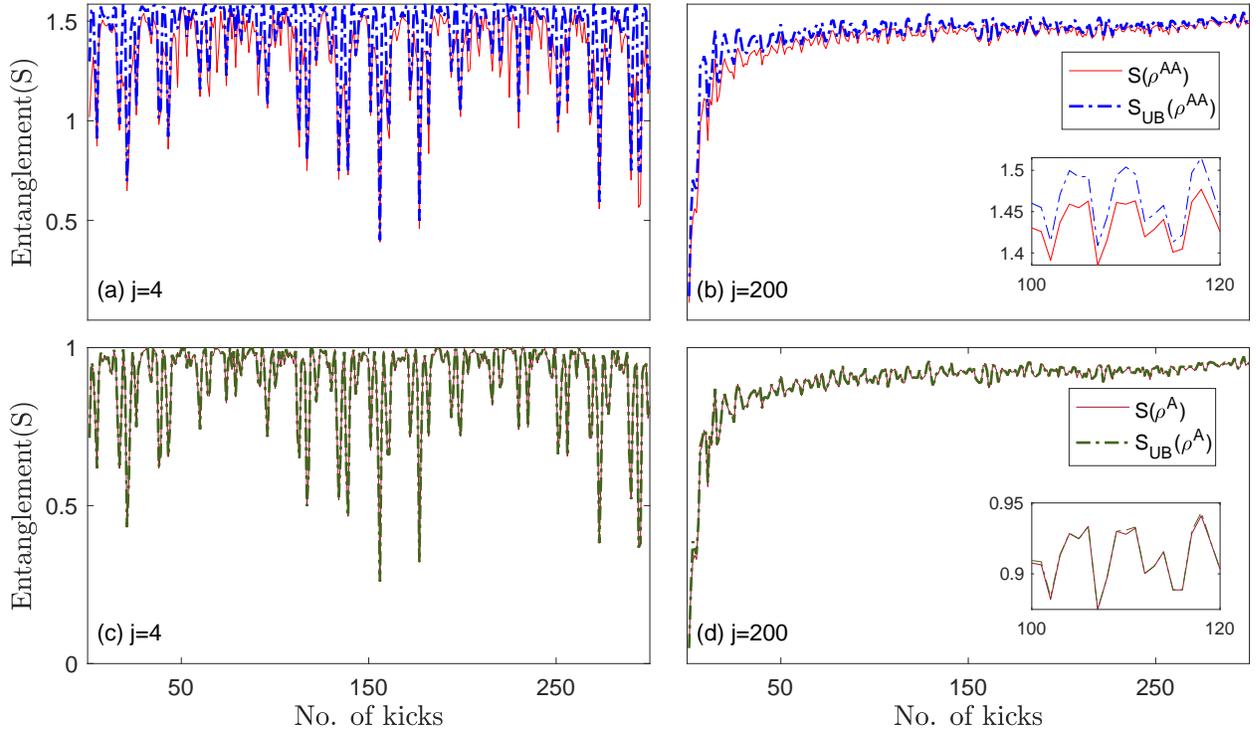}
\caption{Evolution of QKT entanglement and the upper bound in \eqref{bound2} for an initial state in the chaotic region, $P_2$ in Fig. \ref{Classical}. (a) $2{:}(N-2)$ entanglement for $j=4$, (b) $2{:}(N-2)$ entanglement for $j=200$, (c) $1{:}(N-1)$ entanglement for $j=4$, and (d) $1{:}(N-1)$ entanglement for $j=200$. The upper bound on entanglement is almost saturated by the entanglement generated in the kicked top.} 
\label{Chaotic}
\end{figure*}

\begin{figure}
\centering\includegraphics[width=0.49\textwidth]{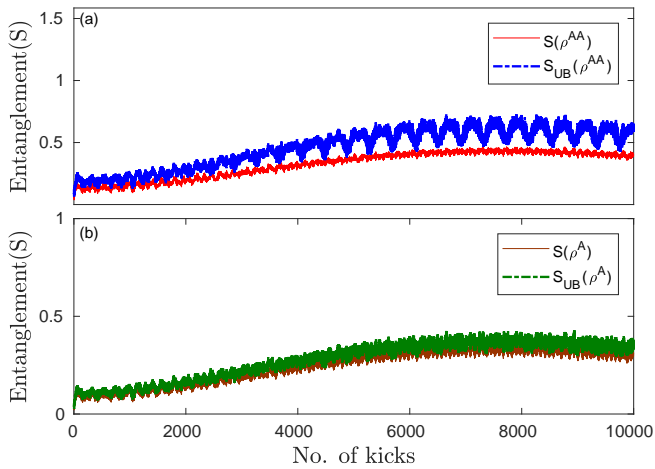}
\caption{Long term evolution of QKT entanglement and the upper bound in \eqref{bound2} for an initial state in the regular region, $P_3$ in Fig. \ref{Classical} for $j=100$. (a) $2{:}(N-2)$ entanglement, and (b) $1{:}(N-1)$ entanglement.} 
\label{LongTime}
\end{figure}

We study $1{:}(N-1)$ and $2{:}(N-2)$ partition entanglement in the kicked top (where $N=2j$), measured by $S(\rho^A)$ and $S(\rho^{AA})$ respectively. One and two-qubit subsystems are the most relevant systems for quantum computing protocols and have open questions that we have addressed within our framework. The two-qubit reduced state of the kicked top, $\rho^{AA}$, lies in the $j=1$ symmetric subspace and thus has dimension $d=3$ instead of 4. We consider the evolution of an initial SCS and compute $S(\rho^{A})$, $S(\rho^{AA})$, and the corresponding upper bounds using \eqref{bound2}. We study this in the deep quantum regime as well as in the  semiclassical regime for an initial state in the regular region (point $P_1$ in Fig. \ref{Classical}) and an initial state in the chaotic region (point $P_2$ in Fig. \ref{Classical}). Figure \ref{Regular} shows the evolution for $P_1$, and Fig. \ref{Chaotic} shows the evolution for $P_2$. In all figures, we observe that the entanglement in the QKT almost saturates the upper bound on entanglement calculated using \eqref{bound2}. Thus, our upper bound provides a very good estimate for the $1{:}(N-1)$ and $2{:}(N-2)$ partition entanglement in the QKT both in the deep quantum regime as well as the semiclassical regime irrespective of the underlying classical behavior (regular or chaotic). The long-time evolution (Fig. \ref{LongTime}) results in a small deviation between the upper bound and the actual entanglement. Nevertheless, the deviation is not large, and the upper bound remains a good estimate of the entanglement.  

We now explore the connections between chaos and entanglement in periodically driven (Floquet) spin systems. Though several previous studies have explored whether entanglement exhibits signatures of chaos, there is a still a lack of consensus about this question, even in a specific model such as the QKT. In this model, entanglement was shown to display signatures of chaos in numerical and experimental studies even in a very deep quantum regime such as $j=3/2,3 \text{ and, } 4$ \cite{entanglement2004PRE,Chaos2008,chaudhury2009nature,neill2016ergodic}. While these studies showed higher time-averaged entanglement in chaotic regions compared to regular regions, Lombardi \textit{et. al} \cite{lombardi2011,lombardi2015} found  instances of initial states in regular regions that also led to generation of high entanglement. On the other hand, Ruebeck et. al \cite{pattanayak2017} correlated classical regular dynamics with low or high time-averaged entanglement, and chaos with medium level of entanglement in the deepest possible quantum regime of the QKT, $j=1$. 

Here, we resolve these seemingly conflicting studies by considering the divergence of the quantum states from SCSs in regular versus chaotic regions. Our goal is to clarify the connection between entanglement and chaos by explaining the dynamics in the regular regions and then comparing it to the more well-studied chaotic case, in both quantum and semiclassical regimes.
Our analysis shows that in the deep quantum regime where $j$ is small, the upper bound on the entanglement can be large in both regular and chaotic regions, whereas in a semiclassical regime with higher $j$, the bound on entanglement remains lower in regular regions near stable periodic orbits than in chaotic regions (see Figs. 2 and 3). We first discuss the deep quantum regime where $j$ is very small. In previous work \cite{kumari2018quantum}, we presented criteria for determining the magnitude of the quantum number $j$ at which quantum-classical correspondence will be observed near classical periodic orbits. The criteria require that the SCSs centered on all the points in a periodic orbit be almost orthogonal to each other (overlap of order roughly less than $10^{-10}$) in order to observe a correspondence between the classical and the quantum dynamics near the periodic orbits  on time scales sufficiently long compared to the dynamics. When these criteria are satisfied near stable classical periodic orbits, the evolved states remain close to the SCSs, and thus the bound on entanglement in \eqref{bound2} remains small. However, in a deep quantum regime, our criteria for quantum-classical correspondence are typically violated. The states diverge from the SCSs and this results in a higher upper bound on entanglement in \eqref{bound2} in both regular and chaotic regions. For example, the period-4 orbit, $P4$, violates the criteria for $j \lesssim 20$ \cite{kumari2018quantum}, where $P4$ is $(1,0,0) \rightarrow (0,0,-1) \rightarrow (-1,0,0) \rightarrow (0,0,1) \rightarrow (1,0,0)$  [$(\cdot,\cdot,\cdot)$ refers to the classical coordinates $(J_x/j,J_y/j,J_z/j)$]. Thus the states do not remain close to SCSs and entanglement can be large even close to the regular periodic orbit. In \cite{pattanayak2017}, the high entanglement regions identified in the QKT for $j=1$ are precisely the regions in the vicinity of the period-4 orbit. Our discussion above explains why high entanglement in these regular regions of the kicked top was observed in a deep quantum regime in~\cite{pattanayak2017}.  

\begin{figure}
\centering\includegraphics[width=0.49\textwidth]{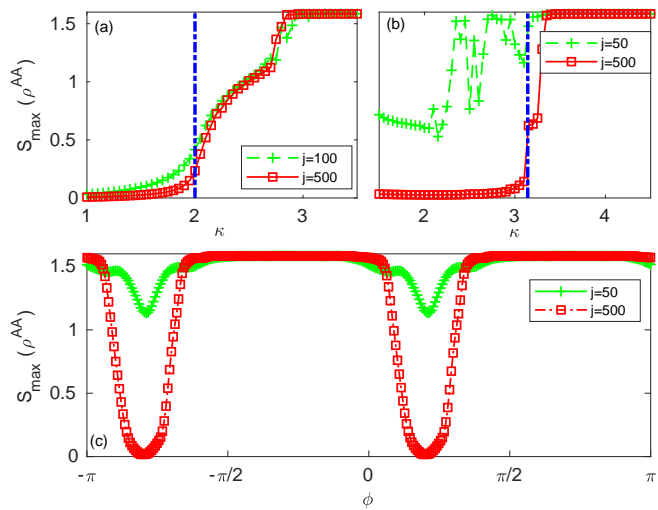}
\caption{Maximum $2{:}(N-2)$ partition entanglement over 5000 kicks in kicked top (a) as a function of $\kappa$ where the initial state is centered at fixed point, $FP_1$, which loses stability at $\kappa=2$, (b) as a function of $\kappa$ where the initial state is centered at a period-4 orbit, $P4$, which loses stability at $\kappa=\pi$, and (c) as a function of initial condition $(\theta,\phi)$ where $\theta=2.25$ is fixed and $\phi$ is varied, $\kappa=3$. The vertical dotted dashed lines in panels (a) and (b) depicts the $\kappa$ value at which bifurcation leads to loss of stability of these orbits.}
\label{MaxEntanglement}
\end{figure}

While large entanglement can be generated in both regular and chaotic regions in the deep quantum regime, the situation changes in the semiclassical regime of large $j$. In general, a regular region in Floquet systems consists of stable periodic orbits, while chaotic regions emerge around the unstable periodic orbits. The criteria for quantum and classical-like states to remain close are satisfied in regular regions but not in chaotic regions in the semiclassical regime. In the QKT, consider the fixed point, $FP_1$ $(0,1,0)$, and the period-4 orbit, $P4$, which lose stability at $\kappa=2$ and $\pi$ respectively \cite{kumari2018quantum}. Figures \ref{MaxEntanglement}(a) and \ref{MaxEntanglement}(b) show that when these orbits are stable, they exhibit very low entanglement for high $j$ values corresponding to the semiclassical limit. In Fig. \ref{MaxEntanglement}(c), the maximum $2{:}(N-2)$ entanglement over 5000 kicks is plotted for a range of initial conditions $(\theta,\phi)$. The maximum entanglement remains very low in regular regions for $j=500$ but not for $j=50$, while chaotic regions always exhibit high entanglement. The same characteristics are seen in Figs. \ref{Regular} and \ref{Chaotic}. 

\section{Summary and Discussion} 
We have presented a general framework to obtain an upper bound on the entanglement in any bipartition of a spin system with a constant angular momentum value $j$. Our framework shows that entanglement generation is associated with a divergence from minimum-uncertainty classical-like spin coherent states (SCSs) as measured by the trace distance. We illustrated in the quantum kicked top (QKT) model that our upper bound estimates the entanglement extremely well in regular as well as chaotic regions for $j$ values in the deep quantum regime as well as the semiclassical regime. This demonstrates that the magnitude of entanglement generation can be inferred from the trace distance between the evolved state and the SCSs. This trace distance, in turn, can be inferred from the localized versus delocalized evolution of the system in the Husimi phase space. Our criteria in Ref. \cite{kumari2018quantum} for quantum-classical correspondence can be used to determine the quantum numbers for which the evolution is localized versus delocalized. Thus, our framework combined with the aforementioned criteria \cite{kumari2018quantum} provides a clear and more nuanced understanding of the relationship between entanglement generation and the underlying classical dynamics, compared to previous studies.

By relating entanglement to the trace distance, our work provides insight and intuition about the quantum-classical connection. It shows that entanglement, a purely quantum phenomenon, grows when the quantum evolution diverges from nearby classical evolution as measured by the trace distance. Thus as the distance between quantum and classical evolution grows, quantum properties like entanglement grow as one might expect. Our analysis makes this argument clear and more quantitative, and explains previous seemingly contradicting results about chaos and entanglement.

Our approach and framework has many interesting characteristics. (a) This framework is very useful for systems with mixed phase space, which are in general difficult to deal with. (b) Whereas past studies have often focused on the linear entropy, which is an approximation of the von Neumann entropy to measure entanglement, our bound applies directly to the von Neumann entropy. (c) Our framework can be used to estimate the entanglement for any bipartition of the system. (d) While any pure state of dimension $d$ in place of $\rho_m^{\text{SCS}}$ would provide an upper bound on the entanglement in the relevant bipartition, our chosen SCSs provide a very good estimate of the actual entanglement. These states are chosen without any optimization and with physical motivation from the Ehrenfest correspondence principle. The particular choice of SCS is not critical in chaotic regions since they generate near to maximal entanglement and hence a large upper bound. However, the choice of SCS is significant in regular regions to obtain a tight bound in order to tease out the differences between regular and chaotic behavior. 

Our approach provides a way to efficiently estimate a loose upper bound for entanglement in spin systems and is thus of interest for experiments, where entanglement is challenging to measure. Our results not only provide insights into the fundamental connections between chaos and entanglement, but are also relevant for applications in condensed matter and quantum computing.

\begin{acknowledgments}
M.K.and S.G. acknowledge support from the Natural Sciences and Engineering Research Council of Canada.
\end{acknowledgments}

\end{document}